\documentclass[tradiabstract]{aa} 
\usepackage{graphicx}
\usepackage{txfonts}

\usepackage{natbib}
\bibpunct{(}{)}{;}{a}{}{,} 

\begin{document}
   \title{Synchrotron radio emission in radio-quiet AGNs}

   \subtitle{}

   \author{W. Ishibashi
          \inst{1,2}
          \and
          T. J.-L. Courvoisier
          \inst{1,2}
          }

   \institute{ ISDC Data Centre for Astrophysics, ch. d'Ecogia 16, 1290 Versoix, Switzerland 
    \and Geneva Observatory, Geneva University, ch. des Maillettes 51, 1290 Sauverny, Switzerland \\
    e-mail: Wakiko.Ishibashi@unige.ch, Thierry.Courvoisier@unige.ch 
             }

   \date{Received 14 May 2010/ Accepted 15 October 2010}


   \abstract{
   The basic mechanism responsible for radio emission in radio-loud active galactic nuclei (AGNs) is assumed to be synchrotron radiation. We suggest here that radio emission in radio-quiet objects is also due to synchrotron radiation of particles accelerated in shocks. We consider generic shocks and study the resulting synchrotron properties. 
   We estimate the synchrotron radio luminosity and compare it with the X-ray component produced by inverse Compton emission. We obtain that the radio to X-ray luminosity ratio is much smaller than unity, with values typical of radio-quiet sources. 
The predicted trends on source parameters, black hole mass and accretion rate, may account for the anticorrelation  between radio-loudness and Eddington ratio observed in different AGN samples. 
}
    


   \authorrunning{ W. Ishibashi \and T. J.-L. Courvoisier}
   \titlerunning{Synchrotron radiation in RQ-AGN}

   \maketitle
%

\section{Introduction}

The relative importance of radio emission, compared to the other wavebands, can vary significantly in different AGN classes. 
This leads to the well-known radio-loud/radio-quiet dichotomy. 
The distinction between radio-loud and radio-quiet objects is based on the radio-loudness parameter, $R = F_{r}/F_{o}$, defined as the ratio of monochromatic luminosities at 5 GHz and optical B band at 4400$\dot{A}$.
The resulting distribution of the $R$ parameter in the Palomar Bright Quasar Survey suggests the existence of two distinct classes \citep{K_et_1989}: radio-quiet ($R \sim 0.1-1$) and radio-loud ($R \sim 10-1000$). Radio-loud AGNs are on average $\sim 10^{3}$ times brighter in the radio band compared to radio-quiet sources. The classical boundary between the two populations is formally set at $R \sim 10$ \citep{K_et_1994}. 

An alternative way to quantify the radio-loudness of a source is to compare the radio and X-ray emission components. 
\citet{T_W_2003} introduced a new parameter ($R_{X}$) as a measure of the radio loudness, by taking the ratio of the 5GHz radio luminosity to the 2-10 keV X-ray luminosity, $R_{X} = \frac{\nu L_{\nu} (5 GHz)}{L(2-10 keV)}$. 
It has been shown that $R_{X}$ is well correlated with the classical radio-loudness parameter $R$, with the advantage that the hard X-ray band is not affected by extinction, and thus can also be used for heavily obscured sources \citep{T_W_2003}.
Based on the $R_{X}$ parameter, the boundary between radio-loud and radio-quiet classes is redefined at log$R_{X} = -2.755 \pm 0.015$ \citep{P_et_2007}. \\

The existence of a bimodality in the radio-loudness distribution is still a debated question. 
Radio-loud objects form only a minority of the AGN population, of the order of $\sim 10\%$, whereas the majority of sources is radio-quiet. 
In radio-loud objects, radio emission is believed to originate in a relativistic jet, but the origin of radio emission in radio-quiet objects is not clearly established yet. 

We investigate here whether synchrotron radiation of relativistic electrons gyrating in a magnetic field could be the physical mechanism responsible for radio emission in radio-quiet objects.
This implies that particle acceleration takes place in the central regions of AGNs.   
One of the well-known mechanisms for particle acceleration is acceleration in shocks \citep[e.g.][]{K_1994}. 
Shocks are known to be an efficient mechanism for accelerating particles up to relativistic energies, with energy being drawn from the bulk kinetic energy of the flow itself. This may account for the presence of relativistic particles contributing to the non-thermal component of the AGN continuum. 
In the presence of magnetic fields, somehow accreted with infalling matter, accelerated particles will give rise to synchrotron emission. 
Here we discuss the synchrotron radio emission resulting from generic shocks, independently of the physical origin of the shock itself.  
In particular, we consider the possible role of shocks in the generation of radio emission in radio-quiet AGNs. \\ 

The paper is organized as follows. 
We first introduce generic shocks in Sect. 2.  
We then derive the radio properties of the synchrotron radiation emitted by shock-accelerated electrons (Sect. 3) and compare the resulting radio luminosity with the X-ray luminosity produced by inverse Compton emission (Sect. 4).
Model results are compared with observations in Sect. 5. 
We also discuss aspects related to the relative strength of radio emission in different AGN classes, 
and conclude that synchrotron emission of electrons accelerated in shocks is a likely explanation to understand the radio emission of radio-quiet AGNs (Sect. 6).


\section{Shocks}

The complex dynamics of accreting matter close to the central black hole is likely to induce dissipative processes in the accretion flow and shocks are naturally expected in such an environment. 
Shocks may be expected in widely different contexts, from accretion disc models  
\citep{Blandford_1994}  to more chaotic accretion scenarios \citep{C_T_2005, Paper_1}. 
Here we consider shocks as a mechanism of particle acceleration, and study the resulting synchrotron radio emission, regardless of the details of any specific model. \\

We consider shocks occurring around $\sim$100 Schwarzschild radii from the central black hole. The radial distance is given in units of the Schwarzschild radius: $R = \zeta R_{S}$, and $\zeta_{100} = \zeta/100$. 
We parametrize the black hole mass and the accretion rate with typical values of radio-quiet AGNs, such as Seyfert galaxies: 
the black hole mass is expressed in units of $M_{BH} = M_{8} \times 10^{8} M_{\odot}$, and the accretion rate in units of $\dot{M} = \dot{M}_{0} \times 1 M_{\odot}/\textrm{yr}$.  

We assume that a small fraction of accreted matter undergoes efficient shock acceleration. 
Particles can cross the shock front several times, with an average energy gain. 
The acceleration process is stochastic and the resulting particle distribution function is described by a power law with index $p$. 
In the optically thin regime, the synchrotron emission is characterized by a power law of the form $\nu^{-(p-1)/2}$, whereas the emission is self-absorbed at low frequencies.
The optically thick regime is predicted to form a rising spectrum of the form $\nu^{5/2}$ (synchrotron self-absorption). \\


\section{Synchrotron radiation}

\subsection{Synchrotron radio luminosity}

The synchrotron power emitted by a single particle is given by \citep{R_L_1979}
\begin{equation}
P_{syn} = \frac{4}{3} c \sigma_{T} \gamma^{2} \beta^{2} u_{B} \, , 
\label{eq_Psyn}
\end{equation}
where $u_{B} = \frac{B^{2}}{8 \pi}$ is the magnetic energy density, with the magnetic field $B$ expressed in units of gauss.
The shape of the synchrotron spectrum is characterized by a peak at a critical frequency
\begin{equation}
\nu_{c} \approx \gamma^{2} \nu_{g} \cong 1.8 \times 10^{7} \gamma^{2} B \; \mathrm{Hz} \, , 
\label{nu_c}
\end{equation}
where $\nu_{g} = \frac{e B}{m_{e} c}$ is the gyro-frequency of a non-relativistic electron. 

To obtain the total power emitted by a population of electrons, we integrate Eq.(\ref{eq_Psyn}) over the electron distribution function $N(\gamma)$:
\begin{equation}
P_{tot} = \int P_{syn} N(\gamma) d\gamma \, . 
\end{equation}
Within the shock acceleration scenario, a combination of energy gain and escape probability leads to a power law distribution in energy for the accelerated particles. 
We thus assume a power law distribution in energy for the electron population
\begin{equation}
N(\gamma) d\gamma = N_{0} \gamma^{-p} d\gamma \, ,
\label{e_df}
\end{equation}
where $p$ is the power law index of the electron distribution function.
Introducing upper and lower cut-offs to the electron distribution, the synchrotron luminosity is given by
\begin{equation}
L_{syn} = P_{tot} = \frac{1}{6 \pi} c \sigma_{T} B^{2} N_{0} \int_{\gamma_{min}}^{\gamma_{max}} \gamma^{2-p} d\gamma \, . 
\label{eq_Lsyn}
\end{equation}
From Eq. (\ref{eq_Lsyn}) we see that the integral is dominated by the upper cut-off for power law indices $p < 3$.
Observations of synchrotron sources generally indicate indices in the range $p \sim 2.0-2.4$. 
The power law index depends on the shock properties; in the case of strong shocks $p = 2$, while Monte Carlo simulations of ultra-relativistic shocks yield $p = 2.2-2.3$ \citep[][and references therein]{A_et_2001,G_2002}. 
Thus the power law index is always smaller than 3, and here we adopt  the value of $p = 2.2$.
The synchrotron luminosity (Eq. \ref{eq_Lsyn}) can thus be approximated by
\begin{equation}
L_{syn} \approx \frac{1}{6 \pi} c \sigma_{T} B^{2} N_{0} \frac{\gamma_{max}^{3-p}}{3-p} \, .
\label{LR_th}
\end{equation}
In  order to evaluate $L_{syn}$, we need estimates of the magnetic field strength, the maximal $\gamma$-factors of the electrons, and the normalization of the electron distribution function.  \\

In general, a direct measurement of the magnetic field in the central regions of galaxies is difficult to obtain. 
This is particularly true in the case of radio-quiet sources. In these objects, radio emission is comparatively weak and the infrared component is dominated by thermal emission from dust, thus precise estimates of the magnetic field are not available. 
Indirect estimates can in principle be obtained from observable quantities, such as the measured characteristic frequency, the electron cooling time or parameters of the synchrotron peak. 
\citet{C_et_1988} obtained a quantitative estimate of the magnetic field in 3C 273, during a synchrotron flare.  
The inferred value was of the order of $B \sim 0.7$ G.  
More recently, magnetic fields in the range $B \sim 0.01-0.1$ G have been reported in 5 High Frequency Peaker radio sources, based on VLBA observations \citep{O_DC_2008}. 
In the following, we parametrize the magnetic field to a typical value of $\sim 0.1$ G ($B = B_{0.1} \times 0.1$ G). 

A rough estimate of the maximal $\gamma$-factor of the electrons can be derived from the break observed in the far-infrared spectral energy distributions (SEDs) of radio-quiet AGNs. 
Radio-quiet objects show a characteristic fall-off from strong infrared to weak radio emissions, corresponding to the transition from thermal emission in the infrared to non-thermal synchrotron radiation in the radio domain \citep{B_1994}. 
In radio-quiet sources, the emission dominating the mid/far infrared region is attributed to thermal emission from the dust component, while the synchrotron contribution becomes negligible in this spectral range \citep{P_et_2000}. 
\citet{H_et_2000} analyzed the infrared to millimetre SEDs of 17 Palomar Green quasars and found a decline starting near 100$\mu$m, followed by a steep fall-off towards the mm region. 
From the observed SEDs, we can see that the turnover occurs somewhere in the mm-cm  region. 
Assuming a cut-off frequency of $\sim 10^{10}$Hz, we obtain a corresponding maximal $\gamma$-factor of the order of $\sim$100 ($\gamma \sim \gamma_{100} \times 100$). 
Indeed, electrons with $\gamma$-factors of this order are needed to explain the radio emission in typical synchrotron emitting regions.

In general, only a small fraction of particles is efficiently accelerated in the shocks, reaching   
the required high-$\gamma$ values. 
\citet{C_C_1989} estimated that a few percent of the total number of electrons arriving in the shock region need to be accelerated to relativistic energies to explain the synchrotron radiation in 3C 273. 
We thus normalize the electron distribution function (Eq. \ref{e_df}) as
\begin{equation}
N_{0} = g_{0.05} \times 0.05 \, \langle N_{e} \rangle \, , 
\end{equation} 
where $\langle N_{e} \rangle = \frac{\dot{M} \cdot t_{dyn}}{m_{p}}$ is the average number of electrons present in the region, where $t_{dyn}$ is the time spent by the electrons in the synchrotron active region which is given by the local free-fall time.  \\

Combining the above arguments and inserting numerical values in Eq. (\ref{LR_th}), we obtain an estimate of the synchrotron luminosity 
\begin{equation}
L_{R} \cong 1.1 \times 10^{39} \,  g_{0.05} B_{0.1}^{2} \gamma_{100}^{0.8} \zeta_{100}^{3/2} \dot{M}_{0} M_{8} \, \mathrm{erg/s} \, . 
\label{LR_S}
\end{equation}

We note that the radio luminosity is of the order of  $L_{R} \sim 10^{39}$ erg/s, and is proportional to the black hole mass and accretion rate, $L_{R} \propto \dot{M}M_{BH}$. 


\subsection{Synchrotron self-absorption}

Synchrotron radiation is known to be self-absorbed below a critical transition frequency, producing the inverted spectrum observable at lower frequencies.
The synchrotron self-absorption coefficient for a power law distribution of accelerated particles is derived in \citet{R_L_1979}. 
Following the notation given in \citet{L_B_2008}, the absorption coefficient $\alpha_{\nu}$ can be written as
\begin{equation}
\alpha_{\nu} = \frac{\sqrt{3} q^{3}}{8 \pi m^{2} c^{2}} \left( \frac{3 q}{2 \pi m c} \right)^{p/2} \Gamma_{3} \Gamma_{4} C_{\gamma} B_{\perp}^{(p+2)/2} \nu^{-(p+4)/2} 
\label{abs_coeff}
\end{equation}
where $\Gamma_{3}$ and $\Gamma_{4}$ are as defined in \citet{L_B_2008}, and $C_{\gamma}$ is given by the normalization of the relativistic electron energy distribution. 
The total number density of relativistic electrons is obtained by integrating over the energy range between $\gamma_{min}$ and $\gamma_{max}$
\begin{equation}
n_{0} = \int_{\gamma_{min}}^{\gamma_{max}} C_{\gamma} \gamma^{-p} d \gamma \, , 
\label{eq_integral}
\end{equation}
where $n_{0} = g_{0.05} \times 0.05 \, n_{e}$ is the fraction of accelerated electrons per unit volume. 
Since the power law index of the electron distribution function is always larger than two ($p = 2.2$), the integral in Eq. (\ref{eq_integral})  is dominated by the lower cut-off that we parametrize by $\gamma_{min} \sim \gamma_{5} \times 5$.

The optical depth for synchrotron radiation is given by the absorption coefficient and the typical size ($R \sim 100 R_{S}$) of the shock region
\begin{equation}
\tau_{\nu} = \int \alpha_{\nu} dR \approx \alpha_{\nu} R \, .
\end{equation}
and is of the order of
\begin{equation}
\tau_{\nu} \sim 4.2 \cdot 10^{4} \, \nu_{GHz}^{-3.1} \gamma_{5}^{1.2} B_{0.1}^{2.1} g_{0.05}  \zeta_{100}^{-1/2} \dot{M}_{0} M_{8}^{-1} \, . 
\end{equation} 

We see that the medium is optically thick at a frequency of $\sim$1GHz, and thus
the central core is expected to be self-absorbed and opaque at low frequencies.  \\

The transition frequency, where the optical depth becomes equal to unity ($\tau_{\nu} \sim 1$), is roughly given by
\begin{equation}
\nu_{t} \cong 31 \, \gamma_{5}^{0.4} B_{0.1}^{0.7} g_{0.05}^{0.3} \zeta_{100}^{-0.2} \dot{M}_{0}^{0.3} M_{8}^{-0.3} \, \mathrm{GHz} \, . 
\end{equation} 

The transition from the optically thick to the optically thin regimes is typically located around a few tens of GHz. 
We thus expect a compact, absorbed core in the nuclei of radio-quiet AGNs at $\sim$GHz frequencies.  
However, the precise value of the turnover frequency depends on the source parameters and the shock properties. 
Inhomogeneities in the magnetic field and superpositions of different shock events lead to a broad distribution of transition frequencies, with $\alpha < 2.5$ over a broad frequency domain.
On the other hand, the importance of synchrotron self-absorption steeply declines with increasing frequency, and the source becomes rapidly transparent at higher frequencies.
At high frequencies, we might then be able to directly see the inner optically thin synchrotron emitting region, although thermal emission from the dust component becomes important in radio-quiet sources.


\section{A measure of radio-loudness: the radio to X-ray luminosity ratio}

The physical process at the origin of X-ray emission is assumed to be Compton upscattering of optical/UV photons. 
In standard accretion disc models, the seed luminosity is given by the the disc luminosity, $L_{seed} = \epsilon \dot{M} c^{2}$.
Alternatively, the seed photons can be generated in optically thick shocks, with the average UV luminosity given by $L_{seed} \cong 1.2 \times 10^{44} \eta_{1/3} \zeta_{UV}^{-1} \dot{M}_{0}$ erg/s, in the clumpy accretion scenario \citep{Paper_1}.  
The resulting X-ray variability and spectral properties have been previously discussed \citep{Paper_2, Paper_3}. 
Independently of the origin of the seed photons, here we parametrize the photon luminosity by the typical value of the optical/UV luminosity, $L_{seed} = L_{s} \times 10^{44} \mathrm{erg/s}$, such as observed in Seyfert galaxies.
Assuming an equilibrium situation, the electron temperature is determined by the balance between Coulomb heating and Compton cooling.
The Compton cooling rate in the non-relativistic limit is given by
\begin{equation}
L_{Compton} = \frac{8 \sigma_{T}}{3 m_{e} c} \cdot u_{ph} \cdot E_{e} \, , 
\label{L_Compton}
\end{equation}
where $u_{ph}$ is the energy density of seed photons. 
The Coulomb heating rate is defined as 
\begin{equation}
L_{Coulomb} = \frac{E_{p}}{t_{C}} \, , 
\end{equation}
where $t_{C}$ is the standard Coulomb energy transfer time. 
The average electron energy is obtained by the equilibrium condition ($L_{Coulomb} = L_{Compton}$) 
\begin{equation}
\frac{E_{e}}{m_{e}c^{2}} \cong 0.7 \, f_{1}^{2/7} L_{s}^{-2/7} \zeta_{100}^{1/7} E_{p, MeV}^{4/7} \dot{M}_{0}^{2/7} \, . 
\end{equation}
The resulting electron temperature is of a few hundred keV, and the Compton cooling of the electrons gives rise to X-ray emission. 
The average X-ray luminosity is of the order of
\begin{equation}
L_{X} \cong 4 \times 10^{43} \, \zeta_{100}^{-5/14} E_{p, MeV}^{4/7} L_{s}^{5/7} \dot{M}_{0}^{9/7} M_{8}^{-1} \, \mathrm{erg/s} \, . 
\label{L_X}
\end{equation}
 
We compute the radio to X-ray luminosity ratio, $L_{R}/L_{X}$, using Eq. (\ref{LR_S}) and the X-ray luminosity calculated in Eq. (\ref{L_X}). 
The resulting $L_{R}/L_{X}$ ratio is given by
\begin{equation}
\frac{L_{R}}{L_{X}} \cong 3 \cdot 10^{-5} \, g_{0.05} B_{0.1}^{2} \gamma_{100}^{0.8} E_{p, MeV}^{-4/7} \zeta_{100}^{13/7} L_{s}^{-5/7} \dot{M}_{0}^{-2/7} M_{8}^{2} \, . 
\end{equation}
We observe that the radio to X-ray luminosity ratio is always much smaller than unity, implying that the synchrotron luminosity is only a small fraction of the inverse Compton luminosity.
The seed photon luminosity is expected to be correlated with the accretion rate. Assuming that the optical/UV luminosity is directly proportional to the accretion rate, the radio to X-ray luminosity ratio scales as

\begin{equation}
\frac{L_{R}}{L_{X}} \propto \left( \frac{\dot{M}}{M_{BH}} \right)^{-1} M_{BH} \, .
\end{equation}

We note that the radio to X-ray luminosity ratio is larger for large black hole mass and/or low Eddington ratio, since $\frac{L}{L_{E}} \propto \frac{\dot{M}}{M_{BH}}$. 


\section{Comparison with observations}

\citet{S_et_2007} analyze the distribution of radio luminosities in different AGN classes and study the dependence of radio-loudness on source parameters, i.e. black hole mass and accretion rate. 
Plotting the radio-loudness parameter versus the Eddington ratio, they find that AGNs form two distinct sequences, termed radio-quiet and radio-loud, respectively. 
The total radio luminosities at 5 GHz ($L_{R} = \nu_{5} \times L_{\nu_{5}}$) measured in Seyfert galaxies and Palomar Green (PG) quasars lie in the range $10^{38} - 10^{40}$erg/s. 
By contrast, the radio luminosities of radio-loud quasars are two to three orders of magnitude higher ($L_{R} > 10^{42}$erg/s).
Thus the radio luminosity we have estimated in Eq. ($\ref{LR_S}$), of the order of $\sim 10^{39}$erg/s, falls within the range observed in radio-quiet sources. \\

The dependence of the radio luminosity on source parameters has been investigated by \citet{L_et_2001}, with particular reference to the FIRST Bright Quasar Survey (FBQS).
These objects have radio luminosities intermediate between radio-quiet and radio-loud classes, and are thought to fill the gap between radio-loud and radio-quiet quasars in the radio luminosity vs. optical luminosity plane. 
The empirical scaling of radio luminosity on black hole mass and accretion rate, obtained by fitting the combined data from the FBQS and PG samples, is of the form:
\begin{equation}
L_{5 GHz} \propto M_{BH}^{1.9 \pm 0.2} (L/L_{E})^{1.0} \, ,
\end{equation} 
at accretion rates of $L/L_{E} \sim 0.1$. 
The Eddington luminosity scales with black hole mass ($L_{E} \propto M_{BH}$) and assuming that the bolometric luminosity is proportional to the accretion rate ($L \propto \dot{M}$), the above empirical relation can be re-written as
\begin{equation}
L_{5 GHz} \propto \dot{M} M_{BH}^{0.9} \, . 
\end{equation}

In our case, the model dependence of the radio luminosity is of the form
\begin{equation}
L_{R} \propto \dot{M} M_{BH} \, ,
\end{equation}
quite similar to the empirical relation found by \citet{L_et_2001}. \\

Concerning the spectral shape of the radio emission, high resolution, multi-frequency, radio measurements for radio-quiet objects are still relatively rare \citep{M_et_2004, U_et_2005} compared to the well-studied radio-loud AGNs. 
 Here we briefly summarize a number of VLA and VLBI radio observations of Seyfert galaxies and radio-quiet quasars.
 A sample of nearby Seyfert galaxies from the Palomar spectroscopic survey has been observed with VLA at two frequencies (6cm and 20 cm); roughly half of these low-luminosity Palomar Seyfert galaxies result to have flat or inverted spectra \citep{H_U_2001, U_H_2001}.
Similarly, \citet{N_et_2000} found that at least 15 of the 18 detected radio cores from a sample of low-luminosity AGNs, observed with high frequency VLA, have flat or inverted spectra. 
More recently, dual VLBI observations of Seyfert galaxies, at 6cm and 18cm, supplemented by literature search have been performed by 
\citet{M_et_2004}.
This results in a majority (81$\%$) having at least one flat or inverted spectrum VLBI component.
The values of the spectral index for Seyfert galaxies range between $-0.3 < \alpha < +2.5$.
\citet{B_et_1996} analysed the radio spectra of radio-quiet quasars with VLA, and concluded that roughly $40\%$ of radio-quiet quasars have flat or inverted spectra cores, quite similar to those observed in radio-loud quasars. 

The flat spectra cores observed in radio-loud sources are usually interpreted in terms of a superposition of distinct synchrotron self-absorbed components becoming opaque at different frequencies \citep{K_P-T_1969}.  
Recent radio variability analysis also suggest that the flat or inverted spectra found in radio-quiet objects are a result of partially opaque synchrotron cores, analogous to those found in the radio-loud counterparts \citep{B_et_2005}.  
Thus the flat spectrum components observed in radio-quiet sources might be indicative of an opaque, self-absorbed central region. 
The observed radio morphology of these objects is usually dominated by an unresolved, compact core, 
with the radio emitting region often confined to the central core. 
Our model thus describes the flat core component observed in Seyfert galaxies and radio-quiet quasars. 
We also note that the presence of absorbed, compact cores in radio-quiet objects seem to be more common than previously thought, as confirmed by the increased detection rate with VLBI-resolution measurements \citep{M_et_2004}.  \\

The relationship between radio and X-ray emission components has been investigated by several authors \citep{T_W_2003, P_et_2007}.  
The relative importance of radio emission is quantified by the $R_{X}$ parameter, which can be compared with our model  $L_{R}/L_{X}$ luminosity ratio.
A mean value of log$R_{X} = -3.64 \pm 0.16$ is reported for a sample of local Seyfert galaxies \citep{P_et_2007}, while a value of $L_{R}/L_{X} \sim 10^{-5}$ is found for the Palomar Green sample, with the X-ray luminosity integrated over the 0.2-20 keV band \citep{L_B_2008}.
Thus the observed radio to X-ray luminosity ratios of radio-quiet AGNs are very small, of the order of $\sim 10^{-4}-10^{-5}$, comparable to our predicted small $L_{R}/L_{X}$ ratios. \\


\section{Discussion and conclusion}

The importance of radio emission compared to the other emission components has direct implications on the radio-loudness distribution and its possible dichotomy. 
\citet{P_et_2007} redefine the boundaries between radio-loud and radio-quiet classes at log$R_{X} = -2.755 \pm 0.015$, based on the $R_{X}$ parameter.
According to this criterion, the predicted $L_{R}/L_{X}$ luminosity ratio corresponds to the radio-quiet category. 
However,  it is observed that even within the radio-quiet sequence, the radio-loudness parameter can span several orders of magnitude between PG quasars and Seyfert/LINER objects \citep{S_et_2007}. 
In our picture, the relative strength of radio emission varies depending on the source parameters. 
The $L_{R}/L_{X}$ luminosity ratio is larger for large black hole mass and low accretion rate. 
This may explain the radio-loudness of low-luminosity AGN (LLAGNs), which are known to be accreting at very low Eddington rates and are found to be both underluminous and radio-loud \citep{H_2002}. 
We may therefore account for the broad range of the radio-loudness covered by radio-quiet  AGNs.

Note however that shocks discussed here cannot account for the much higher radio luminosities observed in radio-loud objects. Indeed the radio-loudness parameter of radio-loud AGNs, hosted by giant ellipticals, are up to $\sim$3 orders of magnitude larger \citep{S_et_2007}. In this latter case, radio emission should have a different origin, most probably related to the presence of a relativistic jet. \\

A trend of increasing radio-loudness with decreasing Eddington ratio has been confirmed in numerous studies \citep{H_2002, S_et_2007, P_et_2007}.
Specifically, \citet{P_et_2007} report an anticorrelation between radio-loudness and Eddington ratio (expressed as $L_{2-10 keV}/L_{E}$) followed by a sample of Seyfert galaxies and low luminosity radio galaxies (LLRG): objects with lower $L_{2-10 keV}/L_{E}$ tend to be more radio loud. 
\citet{S_et_2007} confirm this trend of increasing radio loudness with decreasing Eddington ratio, but in addition note that the trend is followed separately by the radio-quiet and the radio-loud sequences. 
We have seen that the model $L_{R}/L_{X}$ luminosity ratio is larger for smaller values of the $\dot{M}/M_{BH}$ parameter. 
Thus the predicted trends with black hole mass and accretion rate are consistent with observations, and may account, at least qualitatively, for the observed $R-L/L_{E}$ anticorrelation. \\

It is now established that radio-quiet AGNs are not radio-silent, and do emit some radio emission. But, contrary to the radio-loud case, the physical origin of radio emission in radio-quiet objects is still not well understood.
Some authors argue for the presence of scaled-down and less powerful jets, by analogy with the radio jets observed in the radio-loud counterparts \citep{U_et_2005}.
\citet{L_B_2008} raised the alternative possibility of magnetic heating by analogy with coronally active stars.
In our picture, radio emission is produced by synchrotron cooling of relativistic electrons accelerated in the shocks occurring within the accretion flow. 
Shocks are known to be a natural site for particle acceleration, and the associated synchrotron radio emission fits naturally into the general picture. 
The resulting radio to X-ray luminosity ratios are very small, with values typical of radio-quiet sources.
We therefore suggest shocks as a possible explanation for the origin of radio emission in radio-quiet AGNs.



\bibliographystyle{aa}
\bibliography{biblio.bib}

\end{document}